\documentclass[12pt,a4paper,twoside]{article}

\usepackage{authblk}

\usepackage[utf8]{inputenc}
\usepackage[T1]{fontenc}
\usepackage{libertine}
\usepackage{microtype} % Slightly tweak font spacing for aesthetics

\usepackage{mathrsfs}
\usepackage[english]{babel} % Language hyphenation and typographical rules
\usepackage{bbm}
\usepackage{amsmath,amsfonts,amscd,amssymb,latexsym}

\usepackage{graphics}
\graphicspath{{/}}
\DeclareGraphicsExtensions{.pdf,.png,.jpg}

\usepackage{rotating}
\usepackage{longtable}
\usepackage{multirow}
\usepackage{cite}
\usepackage{geometry} % ������ ���� ��������
\usepackage[hang, small,labelfont=bf,up,textfont=up]{caption} % Custom captions under/above floats in tables or figures
\usepackage{booktabs} % Horizontal rules in tables
\usepackage{float} % Required for tables and figures in the multi-column environment - they need to be placed in specific locations with the [H] (e.g. \begin{table}[H])
\usepackage{hyperref} % For hyperlinks in the PDF

\usepackage{lettrine} % The lettrine is the first enlarged letter at the beginning of the text
\usepackage{paralist} % Used for the compactitem environment which makes bullet points with less space between them

\usepackage{abstract} % Allows abstract customization
\usepackage{titlesec} % Allows customization of titles
\titleformat{\section}[block]{\large\scshape}{\thesection.}{1em}{} % Change the look of the section titles
\titleformat{\subsection}[block]{\large}{\thesubsection.}{1em}{} % Change the look of the section titles

\usepackage{fancyhdr} % Headers and footers
\newcommand{\be}{\begin{eqnarray}}
\newcommand{\ee}{\end{eqnarray}}
\title{\fontsize{16pt}{12pt}\selectfont\textbf{Rates of particle production in $R^2$ gravity and supersymmetry-kind dark matter}} % Article title

\author[1,2]{E.V. Arbuzova\thanks{Corresponding Author: arbuzova@uni-dubna.ru}}
\author[2,3]{A.D. Dolgov}
\affil[1]{\small{Dubna State University, Dubna, 141983, Russia
 }}
\affil[2]{Novosibirsk State University, Novosibirsk, 630090, Russia }
\affil[3]{BLTP JINR, Dubna, 141980, Russia}
\date{}
\begin{document}
\maketitle % Insert title
\thispagestyle{fancy} % All pages have headers and footers

%-------------------------------------------------------------------------------
%	ABSTRACT
%---------------------------------------

%-------------------------------------------------------------------------------
%	ABSTRACT
%-------------------------------------------------------------------------------

\begin{abstract}

Universe heating in $R^2$-modified gravity is considered. The rates of particle production by the scalaron are calculated for different 
decay channels. Freezing of massive stable relics with the interaction strength typical for supersymmetry is studied. It is shown that 
the bounds on masses of  supersymmetry-kind particles allowing them to form the cosmological dark matter (DM)
depend upon the dominant decay mode of the scalaron. 
In any case the results presented open much wider mass window for DM with the interaction strength typical for 
supersymmetry.

\begin{flushright}
\textbf{DOI}: 1095.85/inp.2022.01010001\\
\textbf{Received} 01.06.2022\\ 
\textbf{Revised} 15.07.2022\\
\textbf{Accepted} 17.07.2022\\
\end{flushright}
\end{abstract}
\hrule

%-------------------------------------------------------------------------------
%	ARTICLE CONTENTS
%-------------------------------------------------------------------------------

\section{Dark matter mystery}

The mystery of dark matter (DM) is one of the central problems of modern cosmology. Dark matter is 
 an invisible form of matter which  up to the present time discloses itself only
through its gravitational action. 
An accepted property of DM particles is that they are electrically neutral  or maybe millicharged, 
since they do not scatter light 
(hence named {\it dark} matter). Otherwise their properties are practically unknown, so particles of many different types could be DM candidates. 

According to observations the fractional mass density of dark matter is:
\be 
\Omega_{DM} = \frac{\rho_{DM}}{\rho_{crit}} \approx 0.265. 
\label{Om-DM}
\ee
 With the critical energy density of the universe:
  \be 
\rho_{crit} = \frac{3H_0^2 M_{Pl}^2}{8 \pi } \approx 5 \ {\rm keV/cm}^3, \label{rho-crit}
\ee  
where $M_{Pl}=1.22 \cdot 10^{19}$ GeV $=2.18 \cdot 10^{-5}$ g is the Planck mass, 
and with the present day value of the Hubble parameter: 
\be 
{H_0 = 100 h\, \rm{ km\, s}^{-1}\,\rm{Mpc}^{-1} \approx 70 \  \rm{km\, s}^{-1}\,\rm{Mpc}^{-1}},
\label{H0}
\ee    
we can find that the observed mass density of dark matter in contemporary universe is: 
\be 
\rho_{DM} \approx 1 \, {\rm keV/cm}^3.
\label{rho-DM-0}
 \ee 

The existence of dark matter and the magnitude of its contribution into the total mass density of the universe follow 
from the analysis of  several independent pieces of data, which include:
\begin{itemize} 
\item	flat rotational curves around galaxies;
\item equilibrium of hot gas in rich galactic clusters;
\item	 the spectrum of the angular fluctuations of Cosmic Microwave Background (CMB) Radiation;
\item onset of Large Scale Structure (LSS) formation at the redshift $z_{LSS}=10^4$ prior to hydrogen recombination at $z_{rec} = 1100$. 
\end{itemize}

Presently, possible carries of dark matter are supposed to belong to two distinct groups
 consisting of microscopically small (elementary particles) and
macroscopically large objects. 
For DM  in the form of elementary
particles the abbreviation WIMP (Weakly Interacting Massive Particle) is used. This group includes such 
particles as axions with masses about $ 10^{-5}$ eV or even smaller, 
heavy neutral leptons with the masses of  a few GeV, 
particles of mirror matter and many others. Until recently, one of the most popular candidate was the Lightest Supersymmetric
 Particle  (LSP), which we are going to consider in what follows. 
 
 The macroscopic dark matter of stellar size is abbreviated as MACHO (Massive Astrophysical Compact Halo Object) and may include 
 primordial black holes with masses started 
 from $\sim 10^{20}$ g up to tens solar masses, topological or non-topological solitons,
possible macroscopic objects consisting e.g. from the mirror matter, {\it etc.}

According to the low energy supersymmetric model, the LSP should have a mass of several hundred GeV, $M_{LSP} \sim $ 100--1000 GeV. 
However, the absence of signal from supersymmetric partners at Large Hadron Collider (LHC), if not excluded, but considerably 
restricts the parameter space open for the minimal SUSY model.    

The cosmological energy density of LSP is proportional to their mass squared: 
\be 
\rho_{LSP} \sim \rho_{DM}^{(obs)} (M_{LSP}/ 1\,{TeV})^2, 
\label{rho-LSP}
\ee
where $\rho_{DM}^{(obs)} \approx 1$ keV/cm$^3$ is the observed 
value of the cosmological density of DM,  see  Eq. (\ref{rho-DM-0}). 
For LSP with the mass $M_{LSP} \sim 1$ TeV, their energy density,  $\rho_{LSP}$,  would be
of the order of the observed dark matter energy density. For larger masses LSPs would overclose  the universe.
These unfortunate circumstances practically exclude LSP as DM particles in the conventional cosmology. However, 
in $(R+R^2)$-gravity the energy density of LSPs may be much lower 
\cite{Arbuzova:2018ydn,Arbuzova:2018apk,Arbuzova:2020etv} and this fact reopens for LSPs the chance  to be 
the constituents of dark matter, if their mass  $M_{LSP} \geq 10^6$GeV 
(for  a review see \cite{Arbuzova:2021etq}).

\section{Cosmological evolution in $R^2$-gravity}

Theory of gravitational interactions, General Relativity (GR), based on the Einstein-Hilbert action:
\be
{ S_{EH}= - \frac{M_{Pl}^2}{16\pi} \int d^4 x \sqrt{-g}\,R},
\label{S-GR}
\ee 
describes basic properties of the universe in very good agreement with observations. However, some features of the universe request to 
go beyond the frameworks of GR. In theories of modified gravity it is achieved by an addition of a non-linear function of curvature, $F(R)$, into 
the canonical action \eqref{S-GR}.  
 
In 1979  V.~T.~Gurovich and A.~A.~Starobinsky suggested to take $F(R)$ proportional to the curvature squared, $R^2$, for elimination of the 
cosmological singularity \cite{Gurovich:1979xg}. In the subsequent paper by Starobinsky  \cite{Starobinsky:1980te} it was found that 
$R^2$-term leads to exponential cosmological expansion  (Starobinsky inflation). The corresponding action has the form:
\be
S_{tot} = -\frac{M_{Pl}^2}{16\pi} \int d^4 x \sqrt{-g}\,\left[R- \frac{R^2}{6M_R^2}\right] + S_{matt},
\label{action-R2}
\ee
where $M_R$ is a constant parameter with dimension of mass and $S_m$ is the matter action. According to the estimate of 
Ref.~\cite{faulkner} the magnitude of temperature fluctuations of CMB demands
$M_R \approx 3\cdot 10^{13} $ GeV. In our paper \cite{Arbuzova:2018ydn} it was found that $R^2$-term 
creates considerable deviation from the Friedmann cosmology in the post-inflationary epoch and, thereby,  
kinetics of massive species in cosmic plasma and the density of DM particles differ significantly from those in the conventional cosmology.

The modified Einstein equations obtained from the action \eqref{action-R2}  have the form:
\be  R_{\mu\nu} - \frac{1}{2}g_{\mu\nu} R -
 \frac{1}{3M_R^2}\left(R_{\mu\nu}-\frac{1}{4}R g_{v}+g_{\mu\nu} D^2-  D_\mu D_\nu\right)R
 =\frac{8\pi}{M_{Pl}^2}T_{\mu\nu}\,, 
 \label{field_eqs}
\ee
where $D_\mu$ is the covariant derivative, $D^2\equiv g^{\mu\nu} D_\mu D_\nu$ is the covariant D'Alembert operator, and
$T_{\mu\nu}$ is the energy-momentum tensor of matter.  

Taking the trace of Eq.~(\ref{field_eqs}) yields
\be
D^2 R + M_R^2 R = - \frac{8 \pi M_R^2}{M_{Pl}^2} \, T^\mu_\mu.
\label{D2-R}
\ee
The General Relativity  limit can be recovered when $M_R\rightarrow \infty$.
In this case, we expect to obtain the usual algebraic relation between the curvature scalar and the trace of the energy-momentum tensor of matter:
\be
M_{Pl}^2 R_{GR} = - 8\pi T_\mu^\mu\,.
\label{G-limit}
\ee

We assume that the cosmological background is described by the spatially flat 
Friedmann-Lema\^itre-Robertson-Walker 
(FLRW) metric:
\be 
ds^2 = dt^2 - a^2(t) \delta_{ij} dx^i dx^j,
\label{FLRW}
\ee 
where $a(t)$ is the cosmological scale factor and $H = \dot a/a$ is the Hubble parameter at an arbitrary time moment. 

We consider the homogeneous and isotropic matter distribution with the linear equation of state: 
\be
P = w \rho,
\label{eq-state}
\ee
where $\rho$ is the energy density, $P$ is the pressure of matter,
and $w$ is usually a constant parameter. For non-relativistic matter 
$w=0$, for relativistic matter $w=1/3$, and for the vacuum-like state $w=-1$. 

Correspondingly, the energy-momentum tensor of matter $T^\mu_ \nu$ has the following  diagonal form:
\be
T^\mu_\nu = diag(\rho, -P, -P, -P).
\label{T-mn}
\ee

For homogeneous field, ${R=R(t)}$, and with equation of state \eqref{eq-state} the evolution of curvature is governed by the 
equation:
\be 
\ddot R + { 3H\dot R} +M_R^2R = - \frac{8\pi M_R^2}{M_{Pl}^2}(1 - 3w)\rho. 
\label{ddot-R}
\ee

In metric (\ref{FLRW}) the curvature scalar is expressed through the Hubble parameter as:
\be
R=-6\dot H-12H^2\,.
\label{R-of-H}
\ee

The energy-momentum tensor satisfies the covariant conservation condition $D_\mu T^\mu_\nu = 0$, which in 
FLRW-metric (\ref{FLRW})  has the form: 
\be
\dot\rho = -3H(\rho+P)  = -3H (1+w) \rho\,.
\label{dot-rho}
\ee

Equation \eqref{ddot-R} for the curvature evolution does not include the effects of particle production by the curvature 
scalar. It  is a good approximation at inflationary epoch, when particle production by $R(t)$ is practically
absent, because $R$ is large  by the absolute value
and  the Hubble
friction is large, so  $R$ very slowly evolves down to zero. At some stage,
when $ H$ becomes smaller than $ M_R$, curvature starts to oscillate  around zero
efficiently producing particles. 
It commemorates the end of inflation, the onset of the heating of the universe, 
which was originally void of matter.
At that moment the transition from the accelerated expansion (inflation) to a de-accelerated one took place.
 The universe evolution is similar to
Friedmann matter dominated regime but still differs  from it in many essential
features.

 As we see in what follows, curvature, $R(t),$ can be considered as an effective scalar field (scalaron) with the 
mass equal to $ M_R$ and with the decay width $ \Gamma $, since the right hand side (r.h.s.) of  
equation of motion of $R(t)$ (\ref{ddot-R-Gam})
exactly coincides with the r.h.s. of the Klein-Gordon equation
for an unstable massive scalar particle.

It is convenient to introduce dimensionless time variable and dimensionless functions:
\be
 \tau =  M_R\,t,\ \ \ H = M_R\, h, \ \ \ R = M_R^2\, r, \ \ \ \rho = M_R^4\, y. \ \ \
\label{dim-less}
\ee
Equations (\ref{ddot-R}), (\ref{R-of-H}),  and (\ref{dot-rho}) now become:
\be
&&r'' + 3h  r' + r = - 8 \pi \mu^2 (1-3w) y, \label{r-two-prime}\\
&&h' + 2h^2 = - r/6, \label{h-prime} \\
&&y' + 3(1+w)h\,y = 0, \label{y-prime}
\ee
where prime means derivative over $\tau$ and $\mu = M_R/M_{Pl}$.

Let us consider first the inflationary stage of "empty" universe with $\rho =0$ ($y=0$ in dimensionless quantities).  

It is known that with sufficiently large initial  absolute value of curvature, $R$, the 
{devoid} of matter universe would expand quasi exponentially \cite{Starobinsky:1980te}  long enough 
to provide solution of flatness, horizon and homogeneity problems existing in Friedmann cosmology (for the review see e.g. the book \cite{Linde:2005ht}).

The scale factor at inflationary stage behaves as: 
\be
a(t) \sim \exp \left[  { \int_0^{t} h(t')\,dt'} \right],   
\label{a-inf}
\ee
  where $a_{inf} = \exp [N_e] $ and 
$N_e$ is the number of e-foldings  achieved
during inflation. The initial conditions should be chosen in such a way that 
at least 70 e-foldings are ensured:
\be
N_e = \int_0^{\tau_{inf}} h\,d\tau \geq 70,
\label{h-dt}
\ee
where $\tau_{inf}$ is the moment when inflation terminated.  This can be  realised if the initial value of $r$
is sufficiently large  and practically independent on the initial value of $h$. 

We can roughly estimate the duration of inflation neglecting higher derivatives in 
Eqs.~(\ref{r-two-prime}) and taking $y=0$, so we arrive to the  following simplified set of equations:
\be
h^2 &=& - r/12, \label{h2}\\
3 h r'  &=& - r  . \label{h-r-prime}
\ee
These equations are solved as:
\be
\sqrt{-r(\tau)} =  \sqrt{-r_0} - \tau/\sqrt 3,
\label{sqrt-r}
\ee
where $r_0$ is the initial value of $r$ at $\tau = 0$. According to Eq. (\ref{h2}), the Hubble parameter  
behaves as $h(\tau)=(\sqrt{-3r_0} - \tau)/6$. The duration of inflation is roughly determined by the condition 
$h=0$:
 \be
 \tau_{inf} = \sqrt{-3r_0}.
 \label{tau-inf}
 \ee 

\begin{figure}[!htbp]
  \centering
  \begin{minipage}[b]{0.4\textwidth}
    \includegraphics[width=\textwidth]{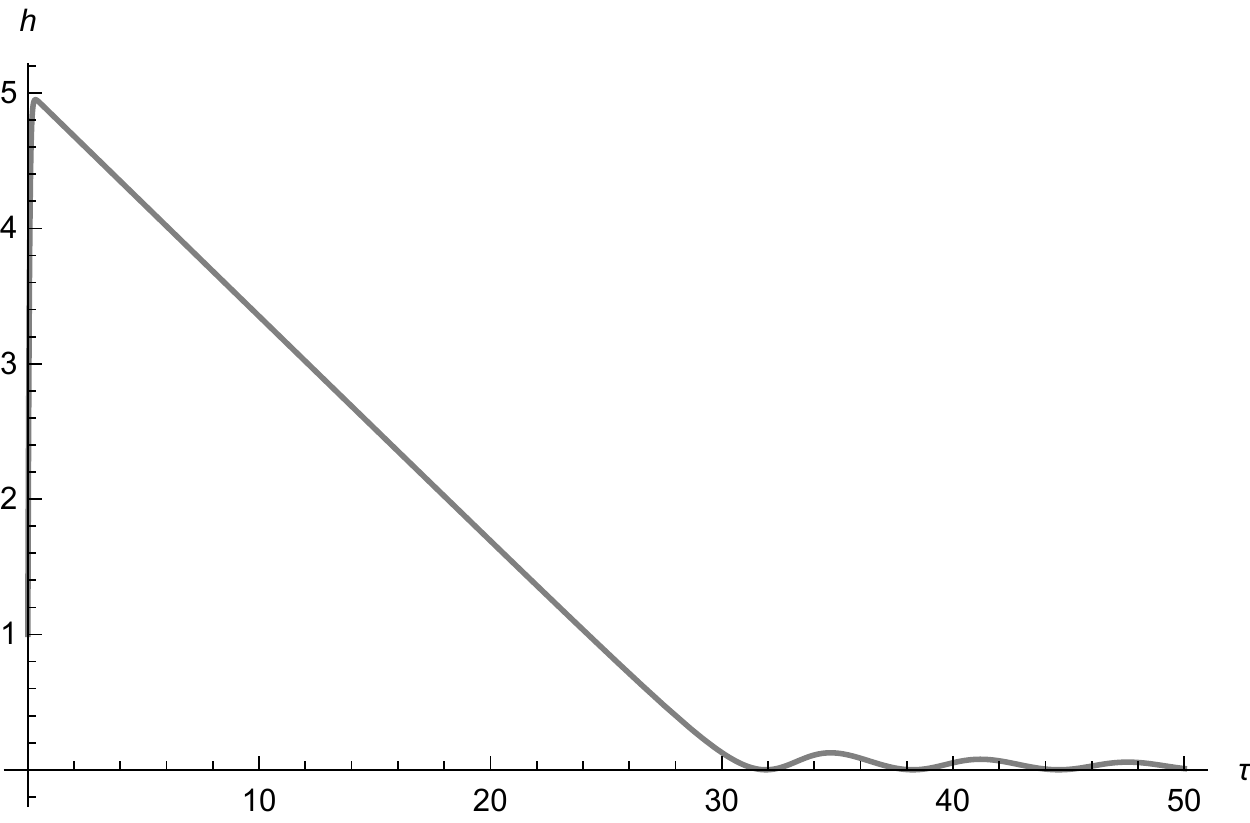}    
  \end{minipage}
  \hspace*{.15cm}
  \begin{minipage}[b]{0.4\textwidth}
    \includegraphics[width=\textwidth]{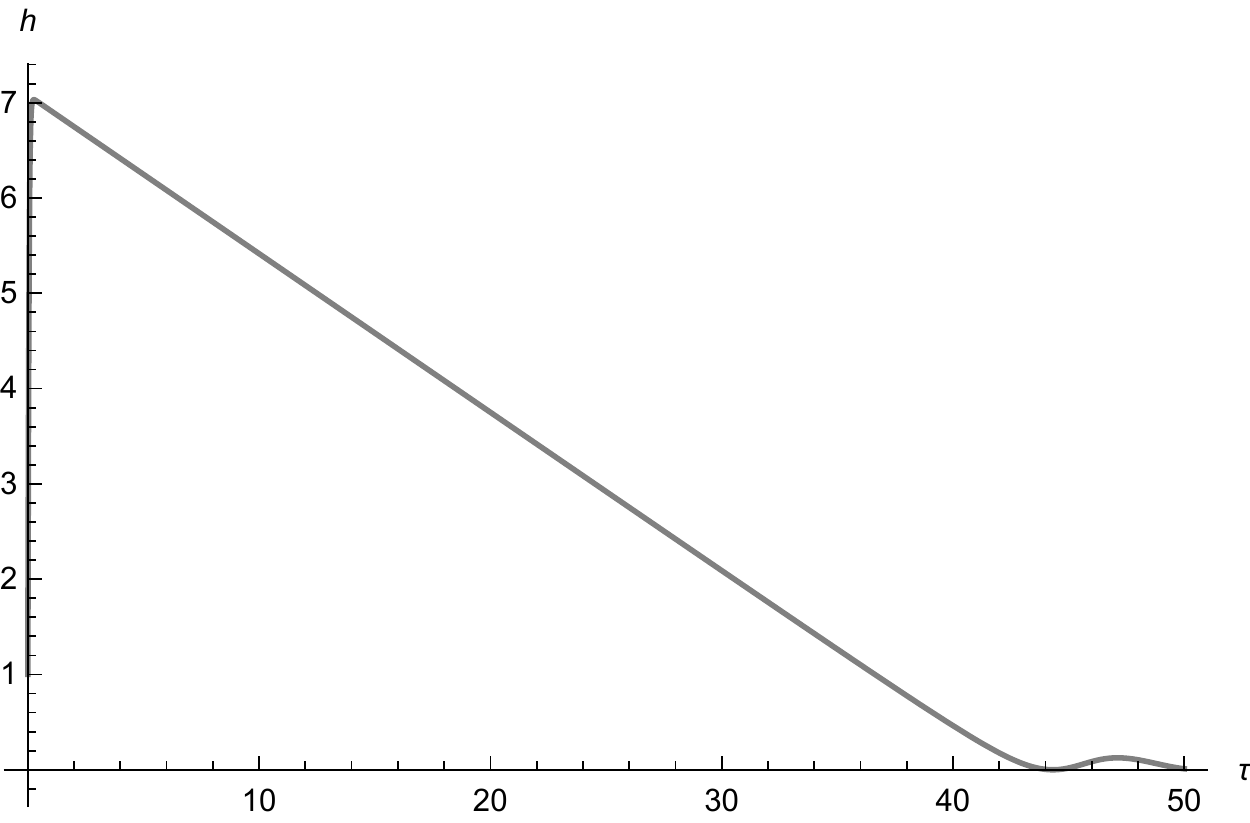}
      \end{minipage}
  \caption{Evolution of $h (\tau)$ at the inflationary stage with the initial values of dimensionless curvature
  $|r_0| = 300$ (left) and $|r_0| = 600$ (right). 
   The numbers of e-foldings, according to  Eq. (\ref{h-dt}), are respectively 75 and 150.}
  \label{f:h-dt}
 \end{figure}

The number of e-folding is equal to the area of the triangle below 
the line $h(\tau)$, thus $N_e \approx |r_0| / 4$. It is in excellent agreement with numerical solutions of
Eqs.~(\ref{r-two-prime})-\eqref{h-prime} depicted in Fig.~\ref{f:h-dt}.
After  the Hubble parameter, $H(t),$ reached 0, it started to oscillate around it with the amplitude decreasing as $2/(3t)$ and 
the exponential rise of a scale factor, $a(t)$, turns into a power law one.  

\begin{figure}[!htbp]
  \centering
    \includegraphics[width=0.45\textwidth]{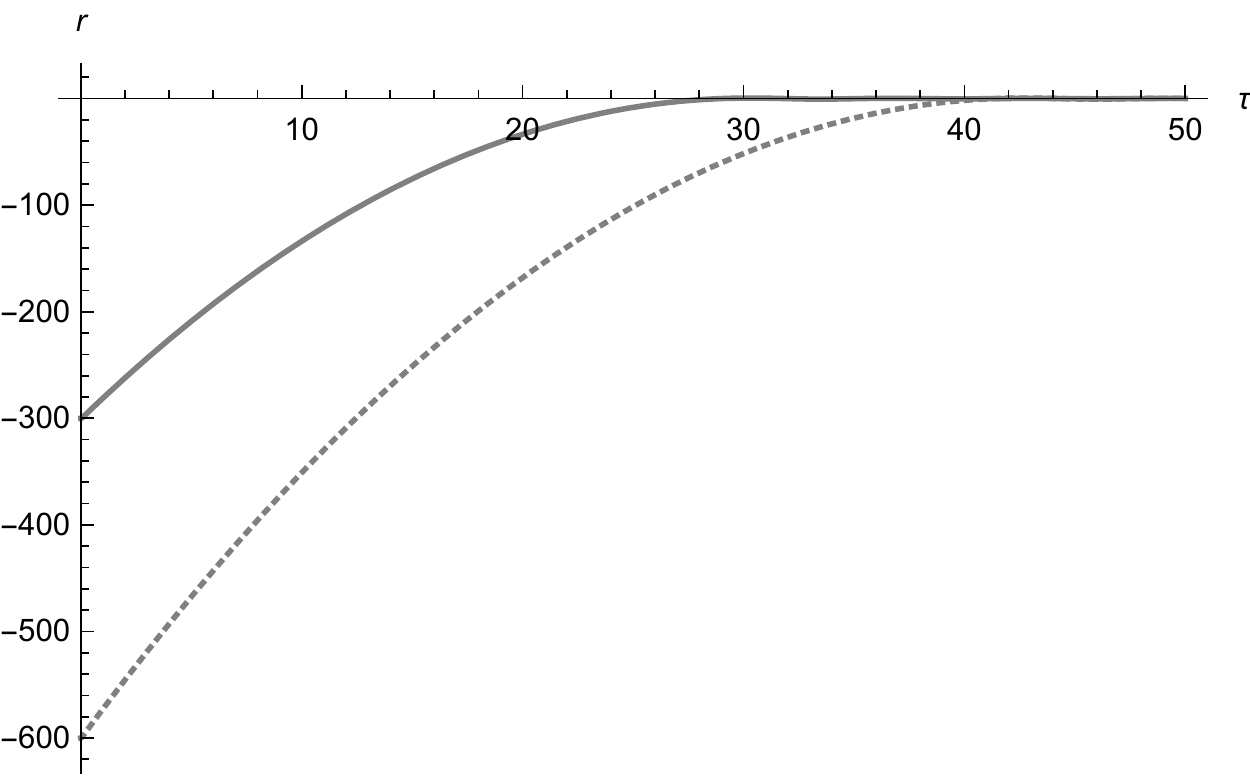}
    \begin{minipage}[b]{0.45\textwidth}
    \includegraphics[width=\textwidth]{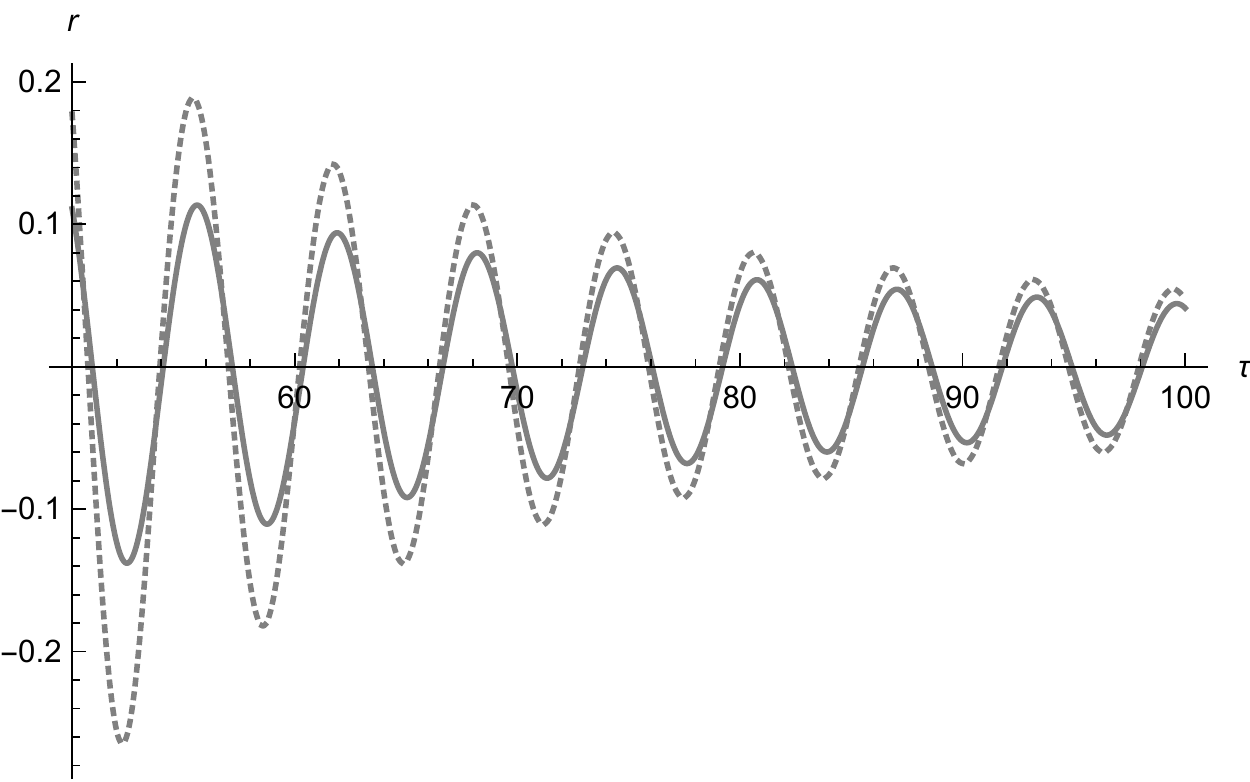}
\end{minipage}
 \caption{Evolution of the dimensionless curvature scalar for $r_0=-300$ (solid)
and  $r_0=-600$ (dotted). 
Left panel: evolution during inflation; right panel: evolution after the end of inflation when curvature scalar starts to oscillate
(scale differs from the left graph).
}
  \label{f:r36}
 \end{figure}

\begin{figure}[!htbp]
  \centering
  \begin{minipage}[b]{0.45\textwidth}
    \includegraphics[width=\textwidth]{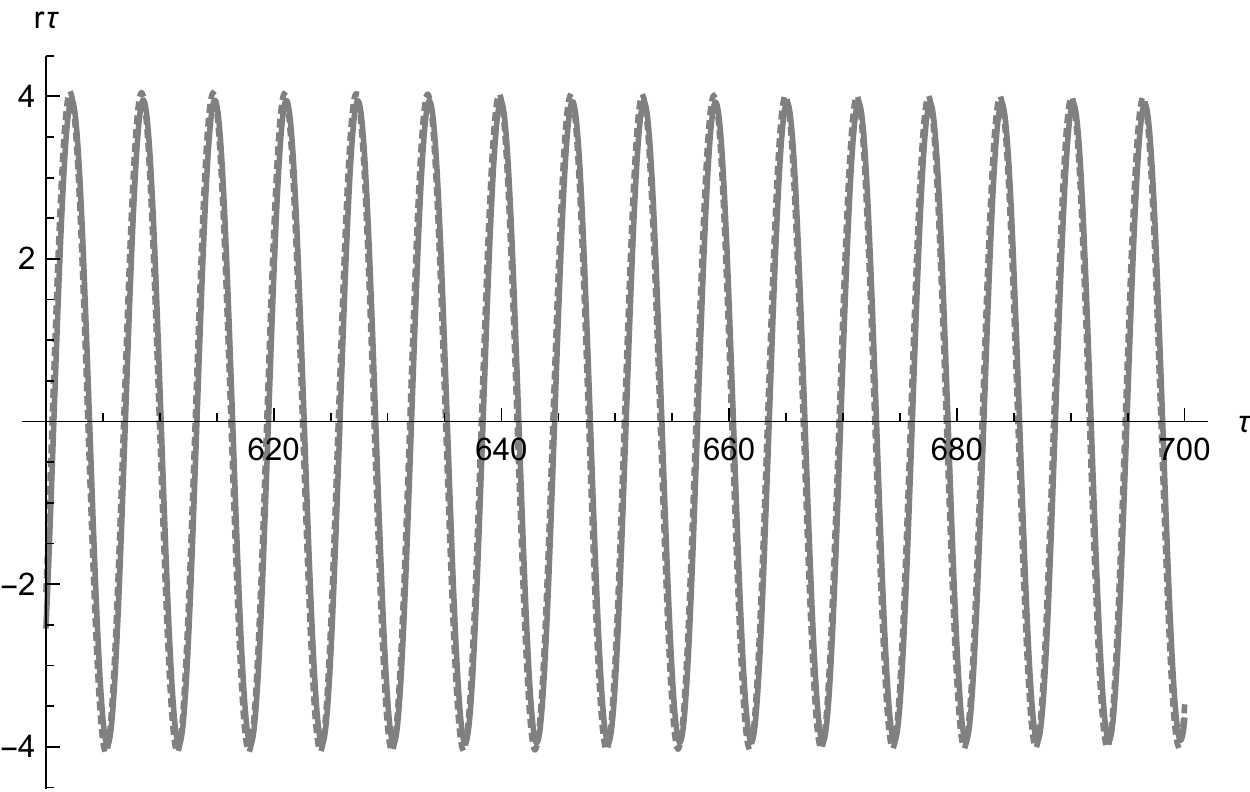}
      \end{minipage}
  \begin{minipage}[b]{0.45\textwidth}
    \includegraphics[width=\textwidth]{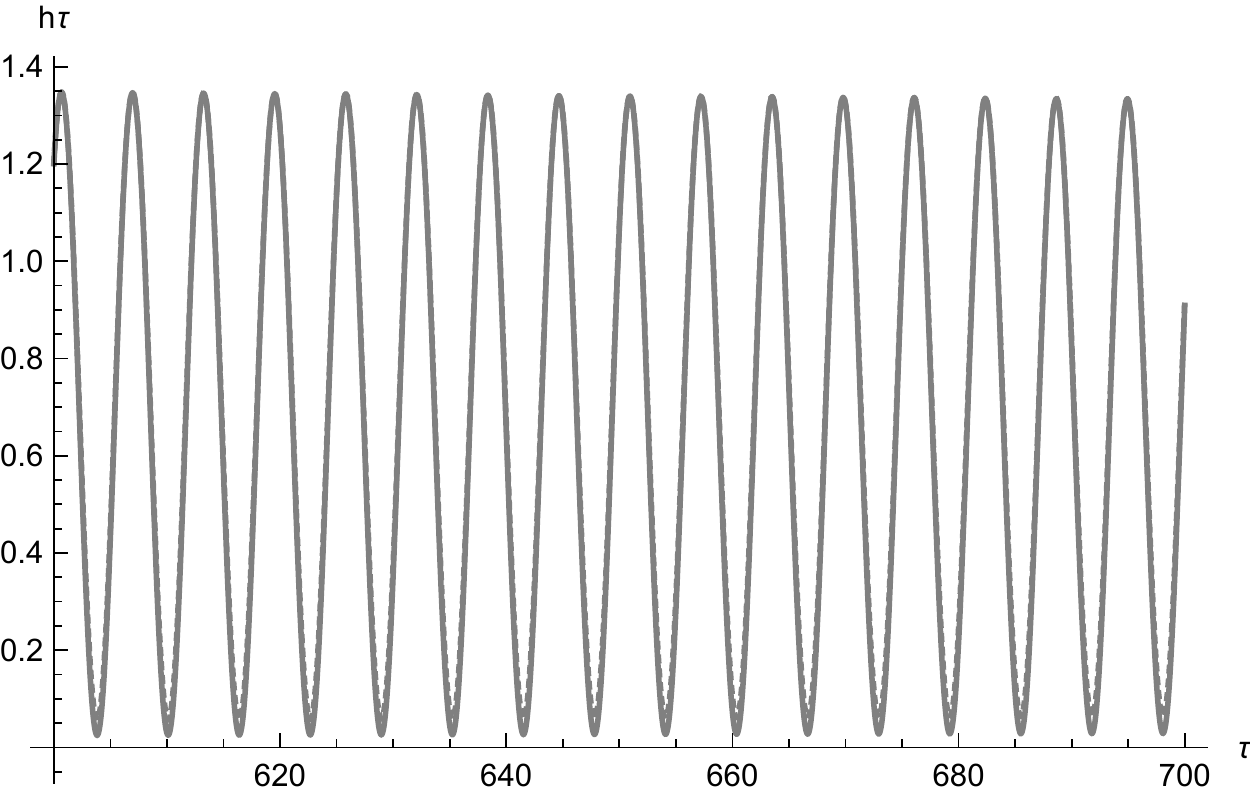}
\end{minipage}
  \caption{Evolution of the curvature scalar $ r(\tau)  \tau$ (left panel) and the Hubble parameter $ h(\tau)  \tau$ (right panel) 
  at post-inflationary epoch as functions of dimensionless time $\tau $. 
 }
  \label{f:r-postinf}
 \end{figure}

The evolution of the dimensionless curvature scalar, $r$, during inflation is presented in the left panel of Fig.~\ref{f:r36}. 
Inflation terminates when both, $h$ and $r$, reach zero. Numerical solution for $r(\tau)$ immediately  after the end of inflation  is presented in the right panel of Fig.~\ref{f:r36}. 
For larger $\tau $ the solutions, $r(\tau)\tau $ and $h(\tau)\tau $, take very 
simple forms depicted in Fig.~\ref{f:r-postinf}. Both $ r(\tau) \tau $ and $h(\tau)\tau $ oscillate with constant amplitudes, so that
the curvature, $\tau r(\tau)$, oscillates 
around zero, while the Hubble parameter, $\tau h(\tau )$, oscillates around $2/3$ 
almost touching zero at the minima.

Stimulated by the numerical solution 
and following Ref.~\cite{Arbuzova:2018ydn}, we find  the asymptotic analytical solutions: 
\be
&&{r=-\frac{4\cos(\tau+ \theta)}{\tau} - \frac{4}{\tau^2}}, \label{rsol} \\
&&h= \frac{2}{3\tau} \left[1+ \sin(\tau + \theta )\right] \label{hsol}, 
\ee
where the constant phase $\theta $ is determined {from}
the initial conditions and 
can be adjusted by the best fit of the asymptotic solution to the numerical one.
Comparison of numerical calculations presented  in Fig.~\ref{f:r-postinf}   
with analytical estimates \eqref{rsol} and \eqref{hsol}
gives ${\theta = -2.9\pi/4}$.

The impact of particle production on the evolution of curvature, $R(t)$, is usually described by an addition of 
the friction term, $\Gamma \dot R$, into the l.h.s. of Eq. (\ref{ddot-R}) 
with a constant~$\Gamma $:
\be 
\ddot R + (3H + { \Gamma}) \dot R+M_R^2R = - \frac{8\pi M_R^2}{M_{Pl}^2}(1 - 3w)\rho
\label{ddot-R-Gam}
\ee
The value of $\Gamma $ depends on the decay channel of the scalaron and is calculated in the section Sec.~\ref{s-decay} for different types of the decay.  

Note, that this simple description of the decay by $\Gamma \dot R$-term in Eq. (\ref{ddot-R-Gam}) is valid only for harmonic oscillations of $R(t)$. For arbitrary temporary evolution of $R$ 
the equation becomes integro-differential, non-local in time 
{one}~\cite{Dolgov:1998wz,Arbuzova:2011fu} and is presented on Sec.~\ref{s-decay}. 

Particle production leads to an emergence of the source term in Eq. \eqref{dot-rho} for the energy density:  
\be 
\dot\rho  + 3H (1+w) \rho  =  \bar S[R] \neq 0.
\label{dot-rho-pp}
\ee
The corresponding system of dimensionless equations takes the following form:
\be 
&& { { h' + 2h^2 =  - r/6, }}\label{h-prime-1} \\ 
&& {{ r'' + (3h + \gamma) r' + r = - 8 \pi \mu^2 (1-3w) y, }}\label{r-two-prime-1}\\  
&& {{y' + 3(1+w)h\,y = S[r], }}\label{y-prime-1}
\ee
where   ${ \mu = M_R/M_{Pl}}$, ${\gamma =  \Gamma/M_R}$. Detailed description of the solution of this system during the 
heating of the universe can be found in our review \cite{Arbuzova:2021etq}. 

According to our calculations the cosmological history in $R^2$-theory can be separated into 4 distinct epochs.   

Firstly, there was the inflationary stage, when the universe was void and dark with slowly 
decreasing curvature scalar, ${R(t)}$. The initial value of $ R$ should be quite large to ensure sufficiently long inflation: 
${R/M_R^2 \gtrsim 10^2}$.

The second epoch began when ${R(t)}$ approached zero and started to oscillate around it periodically changing sign:
\be 
R=-\frac{4M_R\cos(M_Rt+ \theta)}{t}, \, \ \ \  M_R = 3 \cdot 10^{13} \ \text{GeV}.
\label{Rsol}
\ee 
At this stage the Hubble parameter also oscillates almost touching zero: 
\be 
H= \frac{2}{3t} \left[1+ \sin(M_Rt + \theta )\right].  \label{Hsol}
\ee

The curvature oscillations resulted in the onset of creation of usual matter, which remains subdominant. 
We called this period the scalaron dominated regime at which the heating of the universe took place.  
During this time the universe evolution  and particle kinetics were
quite different from  those in General Relativity. The new features of this stage
 open the window for heavy supersymmetry-kind particles to be the cosmological dark matter, modify high temperature baryogenesis, 
 lead to reconsideration of primordial black holes formation, {\it etc.}  

This period was followed by the transition from the scalaron domination to the dominance of the produced matter of 
mostly relativistic particles:
the oscillations of all relevant quantities  damped down exponentially
and the particle production by curvature   became negligible.

Lastly, after complete decay of the scalaron we arrived to the conventional cosmology governed by the General Relativity. 

In the next sections we consider the epoch of { the universe heating}.  In Sec.~\ref{s-decay} we calculate the rates of the production of different
types of particles by the scalaron.  In Sec.~\ref{freezing} we use the obtained decay widths to calculate the freezing of the massive species $ X$ in plasma, 
which was created by the scalaron decays into heavier particles, and find 
the bounds on the masses of $X$-particles, allowing them to form the cosmological dark matter, for different decay modes of the scalaron.

\section{Rates of particle production by the scalaron \label{s-decay}}

The effective action of the scalaron field leading to equation of motion (\ref{D2-R}) can be written as: 
\be 
A_R = \frac{M_{Pl}^2}{48 \pi M_R^4}\,\int d^4x \sqrt{-g} 
\left[ \frac{(DR)^2}{2} - \frac{M_R^2 R^2}{2} - \frac{ 8\pi M_R^2}{M_{Pl}^2}\, T^\mu_\mu \, R \right] .
\label{A-R}
\ee

To determine the energy density of the scalaron field 
we have to redefine this field in such a way that the kinetic term of the new field enters the action 
with the coefficient $1/2$. So the canonically normalized scalar field is~\cite{Arbuzova:2018ydn}:
\be
\Phi = \frac{M_{Pl}}{\sqrt{48 \pi}\, {M_R^2}}\,R, 
\label{Phi}
\ee
where $R(t)$ is given by Eq.~\eqref{Rsol}.

Correspondingly, the energy density of the scalaron field is equal to: 
\be
\rho_R = \rho_\Phi = \frac{\dot\Phi^2 +{M_R^2} \Phi^2}{2} = 
\frac{M_{Pl}^2 (\dot R^2 + {M_R^2} R^2)}{96 \pi {M_R^4}} 
\approx \frac{ M_{Pl}^2}{ 6\pi t^2}.
\label{rho-R-1}
\ee

The actions of the non-interacting, except for coupling to gravity, complex and real
scalar fields with mass $m$ have respectively the forms: 
\be
 S_{c} [\phi_c]&=& \int d^4x\,\sqrt{-g}\,\left(g^{\mu\nu}\partial_\mu\phi_c^*\,\partial_\nu\phi_c - m^2 |\phi_c|^2
 + \xi R |\phi_c |^2\right), 
\label{A-phi-compl} \\
S_r [\phi_r] &=& \frac{1}{2}\int d^4x\,\sqrt{-g}\,\left(g^{\mu\nu}\partial_\mu\phi_r\,\partial_\nu\phi _r
-m^2 \phi_r^2 + \xi R \phi^2_r\right).
\label{A-phi-real}
\ee
If the constant $\xi$ is zero, fields $\phi$'s are called minimally coupled to gravity; for $\xi = 1/6$ 
they are called conformally coupled,
because in this case the trace of the energy-momentum tensor of the fields $\phi$'s become zero.

The equation of motion both for real and complex fields $\phi$'s have the form 
\be
D^2 \phi + m^2 \phi - \xi R \phi = 0,
\label{D2-phi}
\ee
which in FLRW-metric (\ref{FLRW}) transforms to
\be
\ddot\phi - {\frac{\Delta\phi}{a^2}} +
3H\dot\phi + m^{2}\phi - \xi R\,\phi = 0, 
\label{ddot-phi}
\ee
where $\Delta$ is the three-dimensional Laplace operator in flat 3D-space.

The energy-momentum tensor of $\phi$ is defined as the variation of the action over the metric tensor: 
\be
T_{\mu\nu} = \frac{2}{\sqrt{-g}}\, \frac{\delta S}{ \delta g^{\mu\nu}}. 
\label{T-mu-nu}
\ee
Correspondingly for the complex field
\be 
T^{(c)}_{\mu\nu} = && (\partial_\mu \phi_c^*) (\partial_{\nu}\,\phi_c) + (\partial_\nu \phi_c^*) (\partial_{\mu}\,\phi_c)
-g_{\mu\nu} \left(g^{\alpha\beta}\partial_\alpha\phi_c^*\,{\partial_\beta \phi}_c - m^2 |\phi_c|^2\right)\nonumber \\
&& + \xi \left(2R_{\mu\nu} - g_{\mu\nu}\, R\right)\,|\phi_c|^{2} 
 - 2\xi\left(D_{\mu}\,D_\nu\,-g_{\mu\nu}\,D^2\,\right)|\phi_c|^{2},
\label{t-mu-nu-c}
\ee
where $D_\mu$ is the covariant derivative in metric (\ref{FLRW}). The trace of this tensor is:
\be
T^{(c)\,\mu}_{\mu} = 2(6\xi-1) \partial_\mu \phi_c^* \partial^\mu \phi_c + 2\xi (6\xi-1) R |\phi_c|^2
+ 4 (1-3 \xi) m^2 |\phi_c|^2 . 
\label{trace t-c}
\ee
Note that for $\xi = 1/6$ and $m=0$ the trace vanishes.

For the real field $\phi_r$ the energy-momentum tensor has the same form with twice smaller coefficients.

Note, that fields $\phi$'s enter the equation of motion for $R$ (\ref{D2-R}) via the traces of their energy-momentum tensors.

\subsection{Minimally coupled massless scalars mode }

In this section we consider the scalaron decay into massless scalars minimally coupled to gravity. 
The scalaron decay width into two massless (or very low mass) scalar bosons was calculated in
Refs.~\cite{Starobinsky:1980te,Vilenkin_1985,Arbuzova:2011fu}. Here we follow our paper~\cite{Arbuzova:2011fu}, where 
another approach was used based on papers~\cite{Dolgov:1998wz,Dolgov:1994zq}, which allows to derive closed 
equation  in the case of
an arbitrary time evolution of the source field (in our case the scalaron, $R(t)$), while the traditional
methods are valid only for the harmonic oscillations of the source. 

According to Eq.~(\ref{A-phi-compl}) the action for the complex massless scalar field with minimal
coupling to gravity has the form:
\begin{equation}
 S_c^{(00)} [\phi_c] = \int d^4x\,\sqrt{-g}\, g^{\mu\nu}\partial_\mu\phi_c^*\,\partial_\nu\phi_c 
 \label{S-compl-00}
 \end{equation}
and leads to the following equation of motion:
\begin{equation}
 \ddot \phi_c+3H\dot\phi_c-\frac{1}{a^2}\Delta\phi_c=0\,.
\label{d2-phi-00}
\end{equation}

It is convenient to study particle production in terms of the
conformally rescaled field, and the conformal time  defined according to the equations:
\be
\chi_c = a(t)\phi_c, \,\,\, &&d\eta=dt/a(t). 
\label{conf-trans-1}
\ee
The curvature scalar is expressed through the scale factor as
\be 
R = -6 \left( \dot H +2 H^2 \right) = -6 a''/a^3,
\label{R-def}
\ee
here and below prime denotes derivative with respect to conformal time.

The equation of motion for the conformally rescaled field $\chi$ takes the form:
\be
 \chi_c''-\Delta\chi_c+\frac{1}{6}\,a^2R\,\chi_c=0\,,
\label{chi-diprime-1}
\ee
while action (\ref{S-compl-00}) turns into:
\begin{equation}
 S_c^{(00)}[\chi_c] = \int d\eta\,d^3x\,\left(\chi_c'^*\chi_c' - \vec\nabla\chi_c^* \vec\nabla\chi_c 
-\frac{a^2R}{6}|\chi_c |^2\right) .
\label{S-c-chi-00}
\end{equation}

Equation (\ref{D2-R}), which describes the scalaron evolution, can now be written as:
\be
&& R''+2\frac{a'}{a} R'+a^2 M_R^2 R = \\ \nonumber
 &&\frac{16\pi}{a^2}
 \frac{M_R^2}{M_{Pl}^2} \left[ \chi_c'^* \chi_c'-\vec\nabla\chi_c^* \vec\nabla\chi_c +
 \frac{a'^2}{a^2}|\chi_c|^2-\frac{a'}{a}(\chi_c^*\chi_c'+\chi_c'^*\chi_c)\right].
 \label{R-diprime}
 \ee

Our aim is to derive a closed equation for $R$ taking the average value of the $\chi$-dependent quantum 
operators in the r.h.s. of eq.~(\ref{R-diprime}), in presence of classical curvature field $R(\eta)$.
We used the technique of Refs.~\cite{Dolgov:1998wz,Arbuzova:2011fu}, the details of calculations can be found in 
Refs.~\cite{Arbuzova:2021etq,Arbuzova:2021oqa}.  
As a result, we obtained the closed integro-differential equation for the curvature:
\be 
\label{R_with_back_reaction_approx}
\ddot R+3H\dot R+M^2_R R\simeq 
-{\frac{1}{6\pi}}\frac{M_R^2}{M_{Pl}^2}\frac{1}{a^4}
\int_{\eta_0}^\eta d\eta_1\,\frac{a^2(\eta_1)R''(\eta_1)}{\eta-\eta_1} \\ \nonumber
\simeq
-{\frac{1}{6\pi}}\frac{M_R^2}{M_{Pl}^2}\int_{t_0}^t dt_1\,\frac{\ddot R(t_1)}{t-t_1}\,.
\ee
This equation is naturally non-local in time since the effect of particle production depends upon all 
the history of the system evolution. 

Rigorous determination of the decay width of the scalaron is described in Ref.~\cite{Arbuzova:2011fu}. Here we present
it in a simpler and intuitively clear way. We will look for the solution of Eq.~(\ref{R_with_back_reaction_approx})
in the form:
\be 
R = R_{amp} \cos (\omega t + \theta) \exp( -\Gamma t/2),
\label{R-of-gamma}
\ee
where $R_{amp} $ is the slowly varying amplitude of $R$-oscillations, $\theta$ is a constant phase depending 
upon initial conditions, and $\omega$ and $\Gamma$ is to be determined from the equation. 
The term $3 H \dot R$ is not essential in the calculations presented below and will be neglected. 
The exponent is taken equal to $\Gamma t/2$ so the scalaron energy density would 
decrease as $\exp (-\Gamma t)$.

Assuming that $\Gamma$ is small, so the terms of order of $\Gamma^2$ are neglected and treating the r.h.s. of Eq~(\ref{R_with_back_reaction_approx}) 
as perturbation we obtain:
\be
\left[\left(- \omega^2 + M_R^2 \right) \cos (\omega t + \theta) + 
\Gamma \omega \sin(\omega t + \theta)
\right] e^{-\Gamma t/2} = \nonumber\\
{\frac{1}{6\pi}}\frac{\omega^2 M_R^2}{M_{Pl}^2}\, e^{-\Gamma t/2} \int_0^{t-t_0} \frac{d\tau}{\tau} 
\left[\cos(\omega t +\theta) \cos \omega \tau + \sin(\omega t +\theta) \sin \omega \tau \right].
\label{def-Gamma}
\ee
The first, logarithmically divergent, term in the integrand leads to mass renormalization and can be included
into physical $M_R$, while the second term is finite and can be analytically calculated at large upper integration
limit $\omega t$ according to {the well-known integral}
\be
\int_0^\infty \frac{d\tau}{\tau} \sin \omega \tau = \frac{\pi }{2}.
\label{int-exp}
\ee

Comparing the l.h.s. and r.h.s. of Eq. (\ref{def-Gamma}) we can conclude that $\omega = M_R$
and the width of the scalaron decay into a pair of "charged" massless minimally coupled scalars is
\be
{\Gamma_c = \frac{M_R^3}{12M_{Pl}^2}}.
\label{Gamma-c}
\ee

The width of the decay into a pair of neutral identical particles should evidently be twice smaller:
\be {
\Gamma_{r} = \frac{M_R^3}{24 M_{Pl}^2} . }
\label{Gamma-r}
\ee

\subsection{Conformally coupled massive scalars mode  }

The action of a real scalar field $\phi_r$ with the mass $m$ and with non-minimal coupling to gravity, 
{$\xi \phi_r^2 R$}, has the form:
\begin{equation}
\label{phi_action-m}
 S_r^{(m, \xi)}[\phi_r]=\frac{1}{2}\int d^4x\,\sqrt{-g}\,\left(g^{\mu\nu}\partial_\mu\phi_r\,\partial_\nu\phi_r
 + \xi R \phi_r^2 - m^2 \phi_r^2 \right) ,
\end{equation}
leading to the equation of motion:
\begin{equation}
\label{eq_motion_phi-m-2}
 \ddot \phi_r+3H\dot\phi_r-\frac{1}{a^2}\Delta\phi_r +\left(m^2-\xi R\right) \phi_r =0\,.
\end{equation}
We skip the subindex $r$ in what follows.

The particle production will be considered, as above, in terms of the
conformally rescaled field, and the conformal time according to the definitions:
\be
&&\chi = a(t)\phi, \\
&&d\eta=dt/a(t). 
\label{conf-trans}
\ee
Correspondingly equation (\ref{eq_motion_phi-m-2}) transforms into 
\be
\chi''-\Delta\chi + \left(\frac{1}{6} -\xi \right) a^2 R \chi + m_\phi^2 a^2 \chi =0.
\label{chi-diprime}
\ee
Here prime means differentiation over $\eta$ and $ R=-6a''/a^3$. 
The temporal evolution of $R(t)$ is given by Eq.~(\ref{Rsol}).

The particle production by external time-dependent field $V(t)$
was studied in many works dedicated
to the universe heating, see e.g.~\cite{AD-DK}. In particular the production rate was calculated 
perturbatively (but not only) in the book~\cite{Bambi:2015mba} for the (inflaton) field with 
the harmonic dependence on time:
\be
V(\eta) = V_0 \cos (\Omega_c \eta + \theta),
\label{V-of-t}
\ee 
where $V_0 $ as well as $\Omega_c$ may slowly depend on time.
The field $\chi$ satisfies the equation:
\be
\chi'' - \Delta \chi + V(\eta) \chi = 0.
\label{chi-di-prime-gen}
\ee
Since $dt = a d\eta$, and $\Omega dt = \Omega_c d\eta$, the conformal frequency
is $\Omega_c = a \Omega = a M_R$.

It has been shown, see e.g. Eq. (6.40) in the book~\cite{Bambi:2015mba},
that the number density of particles created per unit of conformal time is
\be
n'_{\chi } = \frac{V_0^2}{32 \pi},
\label{dot-N-prod} 
\ee
where the number density is expressed in canonical way through the transformed field
$\chi = a \phi$, namely
\be
n_\chi \sim \chi \partial_\eta \chi \sim a^3 n_\phi.
\label{N-chi-N-phi}
\ee
Hence in physical time we have:
\be
\dot n_\phi = \frac{V_0^2}{32 \pi a^4}, \ \ \ \ \dot \rho_\phi = \frac{1}{2} M_R\, \dot n_\phi,
\label{dot-N-phi}
\ee
where $\rho_\phi$ is the energy density of the produced particles in physical time. 

In the case of conformally coupled decay products, 
(i.e. equation (\ref{chi-diprime}) with 
$\xi = 1/6$ and $m_\phi \neq 0$, but still $m_\phi \ll M_R$, so the phase 
space suppression is not essential),
the interaction leading to the particle production has the form:
\be 
V = m^2 a^2(t) .
\label{V-t}
\ee
Using the solution~ \eqref{Hsol}: 
\be 
H = \frac{\dot a}{a} = \frac{2}{3t} (1 + \sin (M_R t + \theta) ), 
\label{H-of-t-sol}
\ee
we find 
\be 
V = m^2 a^2(t)  \approx m^2 t^{4/3} \exp \left(1 - \frac{4\cos (M_R t + \theta)}{3 t M_R} \right) 
\to  a^2 \frac{4m^2}{3 M_R t} \cos (M_R t + \theta), 
\label{a-of-t}
\ee
 where we have expanded the exponent up to first order in harmonically oscillating term. Correspondingly
the amplitude entering in Eq.~\eqref{dot-N-phi} becomes:
\be
V_0 = \frac{4m^2a^2}{3 M_R t} .
\label{V-0}
\ee 

According to \eqref{dot-N-phi} the energy release from $\phi$-decay into the primeval plasma in this case is:
\be 
\dot\rho_\phi = \frac{1}{2} M_R\, \dot n_\phi = \frac{M_R V_0^2}{64 \pi a^4}
= \frac{m^4}{{36}\pi t^2 M_R}.
\label{dot-rho-0-0}
\ee
The energy density of the scalaron field  is {$\rho_\Phi = M_{Pl}^2 / (6\pi t^2)$} (see Eq. (\ref{rho-R-1})) and hence the
decay width of the scalaron into conformal massive scalars is equal to:
\be 
{ \Gamma(\xi =1/6, m\neq 0)} = \frac{\dot \rho_\phi }{\rho_\Phi} = \frac{m^4}{{36}\pi t^2 M_R} \cdot \frac{6 \pi\, t^2}{M_{Pl}^2} =  
{ \frac{m^4}{6 M_R M_{Pl}^2}}.
\label{Gamma-xi-m}
\ee

\subsection{Massive fermions mode}

Here we calculate the probability of fermion-antifermion pair production by scalar field 
\be
\Phi = \Phi_0 \cos M_R t = \Phi_0 \, \frac{e^{iM_R t} + e^{-iM_R t} } {2},
\label{Phi-cos}
\ee
where canonically normalized field $\Phi $ is connected with the curvature scalar by the relation  \eqref{Phi}. 

The density of fermions produced per unit conformal time  is given by the expression:
\be 
{n'_\psi} = \frac{|g|^2 M_R^2 \Phi_0^2}{16 \pi},
\label{dot-N-psi}
\ee
where $g$ is a coupling constant. 
The relation between the conformally transformed quantities
and the physical ones remains the same as above:
\be
n'_{\tilde \psi} = a^4 \dot n_\psi.
\label{N-prime-of-N-dot}
\ee
In the considered case the particle production is induced by the oscillation of the scale 
factor:
\be
g\Phi_0 \rightarrow  m_\psi a (t) \rightarrow  \frac{2 m_\psi a}{3 t M_R},
\label{m-psi-a}
\ee
compare to Eq. (\ref{a-of-t}). 

Repeating arguments similar to those presented in the previous subsection we find the width of the scalaron 
decay into a pair of fermions with the mass $m_\psi$:
\be
\Gamma_\psi ={\frac{\dot \rho_\psi}{\rho_\Phi}} = \frac{m_\psi^2 M_R}{36\pi t^2} \cdot \frac{6\pi t^2}{M_{Pl}^2} = 
\frac{ m_\psi^2 M_R }{6 M_{Pl}^2}.
\label{N-prime-psi}
\ee

\subsection{Gauge bosons  mode }

Under conformal transformation vector gauge bosons are not transformed, $A_\mu \rightarrow A_\mu$, 
and their equations of motion in terms of conformal time is the same as those in flat 
Minkowski metric.
So in this approximation gauge bosons cannot be created by conformally flat gravitational field. 
This is
truth but not all the truth. Conformal anomaly destroys this conclusion and allows for gauge boson to be
created~\cite{AD-conf-anom}.

Equation of motion of massless gauge field with an account of the anomaly, as derived in
Ref.~\cite{AD-conf-anom}, has the form:
 \be
 A'' - \Delta A + \alpha \kappa \xi'' A' = 0,
 \label{A-di-prime}
 \ee
 where $\alpha$ is the gauge coupling constant squared (for electromagnetic $U(1)$-gauge
 group $\alpha = 1/137$ at low energies),
 $\xi = \ln a$, 
 \be
 \kappa = \frac{11}{3} N - \frac{2}{3} N_F ,
 \label{kappa}
 \ee 
$N$ is the rank of the proper {gauge} group, and $N_F$ is the number of fermion families.

According to the calculations of Ref.~\cite{AD-conf-anom} the number density of the produced
gauge bosons per unit of physical time is
\be
\dot n_g = \frac{\alpha^2 \kappa^2}{{32} \pi} \left( \frac{\ddot a}{a} \right)^2.
\label{dot-N-g}
\ee
Note that 
\be
R = -6 \left( \frac{\ddot a}{a} + \frac{\dot a^2}{a^2} \right) \approx -6 \frac{\ddot a}{a}.
\label{R-of-a}
\ee 
The last approximate equality is valid for quickly oscillating 
$R$ given by Eq.~(\ref{Rsol}).

In Ref.~\cite{AD-conf-anom} this equation was applied to particle production near singularity
in Friedmann cosmology. Here we shall use it for $R^2$-cosmology. To this end one needs
to substitute the average value of $R(t)^2$ taking $\langle \cos^2 (M_R t) \rangle = 1/2$. So
\be
\dot n_g = \frac{\alpha^2 \kappa^2 M_R^2}{144 \pi t^2},
\label{N-g-fin}
\ee
and finally the width of the scalaron decay into two gauge bosons is equal to:
\be
\Gamma_g = {\frac{\dot \rho_g}{\rho_\Phi}} = 
\frac{\alpha^2 \kappa^2 M_R^3}{144 \pi t^2} \cdot \frac{6\pi t^2}{M_{Pl}^2} = 
\frac{ \alpha^2 \kappa^2 M_R^3 }{24 M_{Pl}^2} .
\label{Gamma-g}
\ee

\section{Freezing of massive supersymmetry-kind relics in cosmic plasma \label{freezing}}

In this section we consider the freezing of massive supersymmetry-kind relics in cosmic plasma and obtain the bounds on masses of DM particles, 
following our papers \cite{Arbuzova:2018apk,Arbuzova:2021etq,Arbuzova:2020etv}.

The energy density of the produced particles depends upon the form of their coupling to curvature, $R(t)$, which in turn 
depends upon the dominant decay mode of the scalaron. If scalaron decays into 2 massless scalars minimally coupled to gravity, the 
decay width of the scalaron and the energy density of the produced scalars are correspondingly equal to: 
\be 
\Gamma_s = \frac{M_R^3}{24M_{Pl}^2}, \ \ \ \   \rho_{s} = \frac{M_R^3}{240 \pi t}.
\label{Gamma-s}
\ee
In the case of scalaron decay into a pair of fermions  with mass ${m_f}$ we have:
\be 
\Gamma_f = \frac{ M_R m_f^2 }{6 M_{Pl}^2 }, \ \ \ \ \rho_f = \frac{ M_R m_f^2}{120 \pi t}.
\label{Gamma-f}
\ee
If the scalaron decay is induced by the conformal anomaly then the decay width and the energy density of the produced massless gauge 
bosons are given by the expressions: 
\be
\Gamma_{an} = \frac{\beta_1^2 \alpha^2 N}{96\pi^2}\,\frac{M_R^3}{M_{Pl}^2} , \ \ \ \ 
\rho_{an} = \frac{\beta^2_1 \alpha^2 N}{4 \pi^2} \,\frac{M_R^3}{120 \pi t}, 
\label{Gamma-an}
\ee
where
$\beta_1$ is the first coefficient of the beta-function, $N$ is the rank of the gauge group, 
$ \alpha$ is the gauge coupling constant, which at high energies depends upon the theory.

The presented laws demonstrate much slower decrease of the energy density of matter than normally for relativistic matter at scalaron 
dominated regime, which is ensured by  the influx of energy from the scalaron decay. 

Let us consider the evolution of massive species $X$ in plasma with temperature $T$. 
The number density $n_X$~of particles having mass $M_X$ is governed by the Zeldovich equation~\cite{Zeldovich:1965gev}:
\be 
\dot n_X + 3H n_X = -\langle \sigma_{ann} v \rangle \left( n_X^2 - n^2_{eq} \right), \   
n_{eq} = g_s \left(\frac{M_X T}{2\pi}\right)^{3/2} e^{-M_X/T}, 
\label{dot-n-X}
\ee
where $ \langle \sigma_{ann} v \rangle$ is the thermally averaged annihilation cross-section of X-particles,
$n_{eq}$ is their equilibrium number density, $g_s$ is the number of spin states, $H$ is the Hubble parameter.

If annihilation of non-relativistic particles proceeds in S-wave the thermal averaging is not essential and we have:
\be
\langle \sigma_{ann} v \rangle= \sigma_{ann} v = \frac{\pi \alpha^2 \beta_{ann}}{2M_X^2}. 
\label{S-wave}
\ee
If annihilation proceeds in P-wave and particles are Majorana fermions, the thermally averaged annihilation cross-section 
has the form:
\be 
\langle \sigma_{ann} v \rangle &=&  \frac{3\pi \alpha^2 \beta_{ann}}{2M_X^2} \,\frac{T}{M_X}.
\label{P-wave}
\ee
In Eqs. \eqref{S-wave} -- \eqref{P-wave} $\alpha$ is a coupling constant, in SUSY theories ${\alpha \sim 0.01}$, and 
 ${\beta_{ann}}$ is a numerical parameter proportional to 
the number of annihilation channels, ${\beta_{ann}}\sim 10$. 

We assume that the direct $X$-particle production by curvature ${R(t)}$ 
is suppressed in comparison with the inverse annihilation of light particles into $X\bar X$-pair. 

Some comments are of order here. There are two possible ways to produce $X$-particles: 1) directly through the scalaron decay into a pair $X \bar X$; 
2) by the inverse annihilation of relativistic particles in thermal plasma. 
 Consideration of the direct production of $X \bar X$-pair by scalaron allows to conclude that the energy 
density of the produced particles would be of the order of the observed energy density of dark matter, 
${\rho_X^{(0)} \approx \rho_{DM}}\approx 1 {\rm keV/cm}^3$, if the mass of $X$-particles is ${M_X \approx 10^7}$ GeV. However, 
for such small masses thermal production results in too large ${\rho_{X}}$. For larger masses  $\rho_X^{(0)} $ would be
unacceptably  larger than $\rho_{DM}$.

A possible way out from this "catch-22" is to find a mechanism which suppress the scalaron decay into a pair of  $X \bar X$-particles. And such mechanism 
does exist. Since oscillating curvature scalar creates particles only in symmetric state, the direct production of $X$-particles is forbidden, if they are Majorana fermions, which must be in antisymmetric state.  

We have solved  Zeldovich equation \eqref{dot-n-X} for different decay channels of the scalaron 
and found the frozen (present day)  energy density of the produced $X$-particles. 

For the scalaron decay into massless non-conformal scalars with the decay width \eqref{Gamma-s} the dimensionless Zeldovich equation
has the form:
\be 
{\frac{df}{dx} =  {  - \frac{0.03 g_s \alpha^2 \beta_{ann}}{\pi^3 g_*} \left(\frac{M_R}{M_X}\right)^3} \, \frac{f^2 - f_{eq}^2}{x^5}, \ \  
n_X = n_{in} \left(\frac{a_{in}}{a}\right)^3 { f}, \ \ x=\frac{M_x}{T}}, 
\label{df-dx-R2-2}
\ee
where $f$ is the dimensionless number density of $X$-particles, $x$ is the dimensionless variable, and $g_*$ is the number of relativistic species in plasma. 

Solving Eq. \eqref{df-dx-R2-2} and taking  $g_* \approx 100$, $\alpha \approx 0.01$, $\beta_{ann} \approx 10$, and $M_R = 3\cdot 10^{13} $ GeV we obtain the estimate of the present day energy density of $X$-particles:
\be
\rho_X  
\approx  10^8 \left(\frac{10^{10} {\rm {GeV} }}{M_X} \right) \,\rm{GeV/cm^3}.
\label{rho-X}
\ee
This value is to be compared with the observed energy density of dark matter $\rho_{DM} \approx 1$~keV/cm$^3$.
Formally taken this equation demands that
$X$-particles must have huge mass, much higher than the scalaron mass, $M_R$, to make reasonable dark matter density.
However, if $M_X$ is only 
slightly larger than  $M_R$, the decay of the scalaron into $X\bar X$-channel would be strongly suppressed 
 due to phase space consideration
and such supersymmetry-like particles with the mass slightly larger than $M_R$ could successfully make the cosmological dark matter. 

Let us consider now the scalaron decay into a pair of fermions. The decay width and the energy density of the produced fermions are given 
by Eq.~\eqref{Gamma-f}.  In this case  the decay probability is dominated by the heaviest fermion. 
 In other words, the largest contribution into the cosmological energy density at scalaron dominated regime is presented by the 
 decay into the heaviest fermion species. 
 
 We assume that the mass of the supersymmetry-like particle is considerably smaller than the masses of the other decay products, 
$m_X < m_f$, at least as $m_X \le 0.1 m_f$. Then the direct production of $X$-particles by curvature, $R(t)$,  can  be neglected.  In such a case
$X$-particles are dominantly produced by the secondary reactions in the plasma, which was  created by the scalaron production of heavier particles.

Dimensionless kinetic equation for freezing of fermionic species has the form:
\be 
\frac{df}{dx} =  {- \frac{\alpha ^2 \beta_{ann}}{2 \pi ^3 g_*} \, \frac{ n_{in}\,M_R m_f^2}{m_X^6} }\,  \, \frac{f^2 - f_{eq}^2}{x^5}, 
\label{df-dx-R2-2-f}
\ee
where  $n_{in}=0.09 g_s m_X^3$ is the initial number density of $X$-particles at  $T \sim m_X$. 

The contemporary energy density of $X$-particle can be approximately estimated as 
\be
\rho_X = m_X n_\gamma \left(\frac{ n_X}{ n_{rel}}\right)_h =  7 \cdot 10^{-9} \frac{m_f^3}{m_X M_R}\, {\rm cm^{-3}},
\label{rho-X-f}
\ee
where $n_\gamma \approx 412$/cm$^3$ and we take $\alpha = 0.01$, $\beta_{ann} = 10$, $g_*=100$, $m_f = 10^5$ GeV, and $m_X = 10^4$~GeV. 

This energy density should be close to the energy density of the cosmological dark matter, $\rho_{DM}~\approx~1$ keV/cm$^3$.
It can be easily achieved with $m_X \sim 10^6$ GeV and $m_f \sim 10^7$ GeV:
\be
\rho_ X = 0.23\,  \left(\frac{m_f}{10^7\, {\rm GeV}}\right)^3 \left(\frac{10^6\, {\rm GeV}}{m_X}\right) \, {\rm \frac{keV}{cm^3}}.
\label{rho-fin}
\ee

\begin{figure}[!htbp]
  \centering
 \includegraphics[scale=0.45]{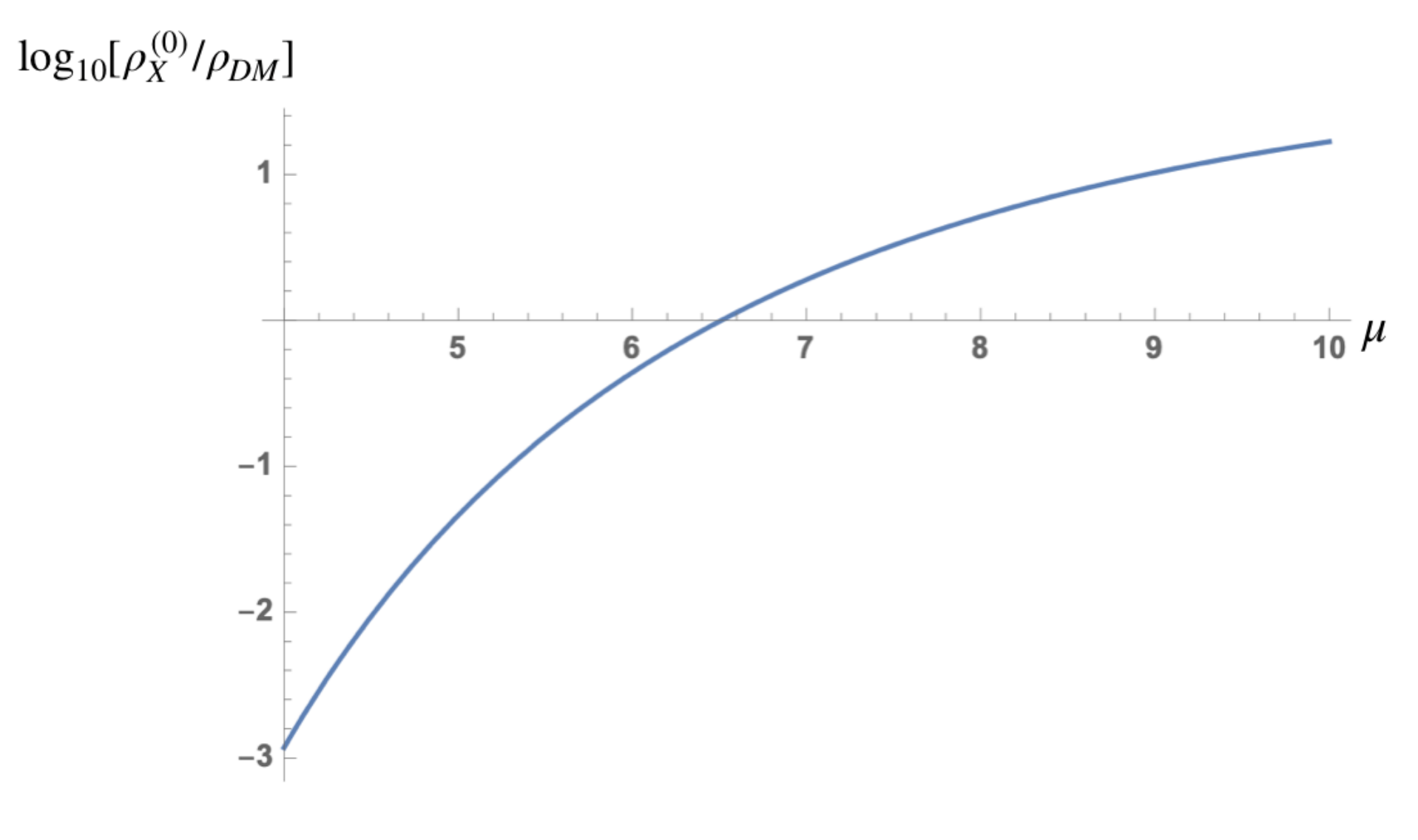}
 \caption{ The logarithm of the ratio of the energy density of $X$-particles with respect to the  observed energy density of dark matter as a function 
  of $\mu =M_R/M_X$.   }
 \label{f:rho-of-MX}
 \end{figure}

In the case when the scalaron decay is induced by the conformal anomaly we use expressions~\eqref{Gamma-an} for the decay width and for the
energy density of the produced gauge bosons and solve the Zeldovich equation \eqref{dot-n-X}. 
The result of calculations of the frozen number density of $X$-particles with mass $M_X$ in cosmic plasma, 
which was created by the scalaron decay into massless gauge bosons due to conformal anomaly, is presented in Fig.~\ref{f:rho-of-MX}. 
 It is clearly seen that if $M_X \approx 5\cdot 10^{12}$ GeV, $X$-particles may be viable candidates for the 
carriers of the cosmological dark matter.

\section{Possible observations \label{s-obs}}

According to results of our papers~\cite{Arbuzova:2018apk,Arbuzova:2018ydn,Arbuzova:2020etv}
the mass of dark matter particles, with the interaction strength typical for supersymmetric ones, can
be in the range from $10^6$ GeV to the value exceeding the scalaron mass,
$M_R =10^{13} $ GeV. It is tempting to find if and how they could be observed,
except for their gravitational effects on galactic and cosmological scales.

There are some interesting possibilities to make $X$-particles visible. The first one is connected with annihilation effects in clusters of dark matter in galaxies and galactic halos, in which, according to Refs.~\cite{Berezinsky:1996eg,Berezinsky:2014wya}, the density of dark matter is 
 much higher than the averaged DM cosmological density. 
Another chance for observation of 
$X$-particles appears if they are unstable. The decay of superheavy DM particles, which could have a lifetime long enough to manifest themselves as stable DM, 
at the same time may lead to the possibly observable contribution to the UHECR spectrum. Heavy $X$-particles would decay though 
formation of virtual black holes, according to  the
Zeldovich mechanism~\cite{zeld-BH-eng,zeld-BH-rus}. 

Below we present simple estimations of the energy flux from the $X$-particle annihilation for the homogeneous dark matter distribution and for more 
realistic case of commonly accepted shape of distribution of dark matter. 

\subsection{Homogeneous dark matter with average cosmological density \label{s-homDM}}
 
 The energy density of dark matter, $\rho_X$, is expressed through the number density,  $n_X$, by the relation $\rho_X = M_X n_X$, where $M_X$ 
 is the  $X$- particle mass. $X$- particle concentration per unit volume is inversely proportional  to their mass and for the observed value of the DM 
 energy density $\rho \approx 1$ keV/cm$^3$ is equal to:
\be
n_X = \frac{1}{M_X} \frac{\text{ keV}}{\text{cm}^3} = \frac{1}{M_{12}} \cdot 10^{-18} \text{cm}^{-3}, 
\ee
where  $M_{12} = {M_X}/({10^{12}}$ {GeV}).

Knowing this value, one can estimate the energy flux of the products of  the annihilation of dark matter particles 
"in the entire Universe" and reaching Earth's detectors. Let us assume that dark matter in the Universe is distributed uniformly and isotropically. 
We assume that the annihilation cross section is determined by the expression
$\sigma v \sim \alpha^2/M_X^2$, where $v$ is the center-of-mass velocity  
and  $\alpha$ is the coupling constant with the typical value $\alpha \sim 10^{-2}$. 
 Then the rate of the decrease of the X-particle density per unit volume is equal to:
\be
\dot n_X = \sigma v n_X^2 = \alpha^2 g_* n_X^2/M_X^2, 
\ee
where $g_*$ is the number of the open annihilation channels, 
$g_* \sim 100$.  

Therefore, the luminosity (energy from the annihilation released per unit time and per unit volume) is equal to:
\be 
\frac{\Delta L}{\Delta V} = \dot \rho_X = E \dot n_X = 2M_X \dot n_X = \frac{2  \alpha^2 g_* n_X^2}{M_X}, 
\ee
where the energy of a pair of annihilating $X$-particles is $E = 2M_X$. 

The energy flux on the Earth originated from the volume element $\Delta V$ at distance $R$ would be:
\be 
\Delta J = \frac{\Delta L}{4 \pi R^2}.
\ee
For a spherical layer with a thickness $\Delta R$ the volume element is:
\be
\Delta V = 4 \pi R^2 \Delta R.
\ee 
Thus, the energy flux from this layer is equal to:
 \be 
\Delta J = \dot \rho_X \Delta R.
\ee
Integrating this expression over $R$ from 0 to $R_{max}$ we obtain the flux of the energy on the Earth from the part of the universe 
contained inside radius $R_{max}$:
\be 
J = \dot \rho_X R_{max} = \frac{2  \alpha^2 g_* n_X^2 R_{max}}{M_X}.
\ee
We take  $R_{max} \approx10^{28}$cm, since above this distance the redshift cutoff is essential, and finally estimate  the energy  flux as:
\be  
J \sim 10^{-39} \frac{\text{GeV}}{\text{cm}^2 \cdot s} \left(\frac{1}{M_{12}}\right)^3.
\ee
The obtained value of the energy flux is negligibly small and cannot be measured with any of the existing detectors.

\subsection{Flux of cosmic rays from DM annihilation in the Galaxy \label{Gal-DM}}

Since the flux of the cosmic rays from DM annihilation is proportional to the square of the DM particle density, smaller objects 
with the number density larger than the average one can create a larger flux of the cosmic rays.  

We take the commonly accepted shape of dark matter distribution \cite{Gunn:1972sv}:   
 \be
 {\displaystyle \rho (r)=\rho _{0}\left[1+\left({\frac {r}{r_{c}}}\right)^{2}\right]^{-1}},
 \label{dm-prof}
 \ee
 where $\rho _{0}$ denotes the finite central density and $r_{c}$ the core radius. We assume for the sake of estimate $r_{c}=1$~kpc and
 calculate $\rho _{0}$ from the condition that at the position of the solar system at $r=8$ kpc:
 \be  
 \rho (8\, \text{kpc})\approx  0.01 M_{\odot} /\text{pc}^3\approx 4 \cdot 10^{-24} \text{g/cm}^3 \approx 2.4\,  \text{GeV/cm}^3.
 \label{rho-solar}
 \ee
Hence we obtain $\rho_0 \approx 150$ GeV/cm$^3$. This value exceeds the average density of cosmological dark matter, 
$\rho_{DM} = 1$~keV/cm$^3$  by 8 orders of magnitude. 

Let us consider the annihilation of DM particles at the point determined by the radius-vector with the spherical coordinates $r, \theta, \phi$ 
directed from the position of the Earth. This point is situated at  the distance $r_b$ from the galactic center:
\be
r_b^2 = r^2 + l^2 - 2 r\,l \cos{\theta},
\label{r-b}
\ee
where $l \approx 8$ kpc is the distance from the Earth to the galactic centre. 

The annihilation creates the flux of cosmic rays per unit time, area, energy, and steradian equal to:
\be
dj= \frac{2 \alpha^2 g_*\,\rho^2(r_b)\, dV}{4 \pi\,M_X^3\,r^2}, 
\ee 
where $r(r_b)$ is determined by Eq. \eqref{dm-prof} and the volume element $dV = r^2 dr d(\cos{\theta}) d\phi $. After integration over $\phi$
we obtain:
\be 
dJ = \frac{ \alpha^2 g_*}{M_X^3}\frac{\rho_0^2\,r_c^4 \, dr\, d(\cos{\theta})}{[r_c^2 + l^2 +r^2  - 2rl \cos\theta ]^2}.
\ee
Integrating over $d\cos\theta $ and $dr$ we obtained the following expression for the total flux:
\be 
J = \frac{ \alpha^2 g_*\,\rho_0^2\,r_c^4 }{M_X^3} \, \int_0^{\infty} \frac{dr}{[(r_c^2 + l^2 +r^2)^2  - 4r^2 l^2 ]}.
\ee

As a result we find the flux of UCHECR at $E=10^{11}$~GeV:
\be 
J_{obs} \sim 10^{-27.6} GeV\,m^{-2}\,s^{-1}\,sr^{-1}.
\label{J-obs}
\ee   

 Another contribution into the
flux of UHECR from possible decays of $X$-particles depends upon their life-time and in particular if 
$X$-particles decay by Zeldovich mechanism through formation of virtual black holes, the contribution to the flux may be essential. 
The precise value is model dependent. 

\section{Results and discussion}

The existence of stable particles with interaction strength typical for supersymmetry and heavier than several TeV 
is in tension with conventional Friedmann cosmology. 
$R^2$-gravity opens a way to save life of such $X$-particles, because in this theory
the density of heavy relics  with respect to the plasma entropy could be noticeably diluted by  radiation from the scalaron decay.
The range of allowed masses of $X$-particles to form cosmological DM depends upon the dominant decay mode of  scalaron.
The results are  summarized in Table~\ref{table:table1}.

\begin{longtable}[]{@{}lll@{}}
\toprule
Dominant decay channel & Decay width & Allowed $M_X$ to form DM\tabularnewline
\midrule
\endhead
Minimally coupled scalars  & $\Gamma_{s} = {M_R^3}/{(24 M_{Pl}^2)}$ & $M_X \gtrsim M_R \approx 3\cdot 10^{13}$ GeV\tabularnewline
Massive fermions & $\Gamma_f = {m_f^2 M_R}/(6 M_{Pl}^2)$ & $M_X \sim 10^6$ GeV\tabularnewline
Gauge bosons & $\Gamma_g = { \alpha^2 \kappa^2 M_R^3 }/(24 M_{Pl}^2)$  & $M_X \sim 5 \cdot 10^{12}$ GeV \tabularnewline
\bottomrule
\caption{The range of allowed masses of $X$-particles to form cosmological DM for different  dominant decay modes of the scalaron.}
\label{table:table1}
\end{longtable}

\section{Conclusion}

Cosmological energy density of dark matter particles is determined by their interaction strength, their mass, and the universe expansion law. 
The frozen density of DM particles is governed by the kinetic Zeldovich equation. In the canonical theory  a former natural 
candidate for dark matter, the lightest supersymmetric particle (LSP), is practically excluded by the recent lower bound on its 
mass obtained at LHC. In this connection it is interesting to consider modified gravity theories which lead to significant deviation
of the expansion regime from that in the conventional cosmology. As a realistic example we discuss $R^2$-inflationary model,
suggested by Starobinsky. We have shown that particle production by oscillating curvature (scalaron) could significantly diminish 
density of stable relics.   
 If the epoch of the domination of the curvature oscillations (scalaron domination) lasted after freezing of massive species, their
density with respect to the plasma entropy could be noticeably diluted by  radiation from the scalaron decay.
Depending on the dominant channel of the scalaron decay the proper value of the mass of the supersymmetry-kind particle, 
forming dark matter, can be in the range from $10^6$ GeV up to $10^{12}$ GeV. The results are presented in Table~\ref{table:table1}. 

\section{Acknowledgments}
The work of E.V. Arbuzova was supported by the RSF grant 22-22-00294, the work of A.D. Dolgov was supported by the RSF grant
20-42-09010.

\end{document}